# Near-Field Microwave Microscopy on nanometer length scales

Atif Imtiaz[1], Marc Pollak, and Steven M Anlage
Center for Superconductivity Research, Department of Physics, University of Maryland, College Park, MD 20742-4111
John D Barry and John Melngailis
Institute for Research in Electronics and Applied Physics, University of Maryland, College Park, MD 20742

**Abstract**

The Near-Field Microwave Microscope (NSMM) can be used to measure ohmic losses of metallic thin films. We report on the presence of a new length scale in the probe-to- sample interaction for the NSMM. We observe that this length scale plays an important role when the tip to sample separation is less than about 10nm. Its origin can be modeled as a tiny protrusion at the end of the tip. The protrusion causes deviation from a logarithmic increase of capacitance versus decreasing height of the probe above the sample. We model this protrusion as a cone at the end of a sphere above an infinite plane. By fitting the frequency shift of the resonator versus height data (which is directly related to capacitance versus height) for our experimental setup, we find the protrusion size to be 3nm to 5nm. For one particular tip, the frequency shift of the NSMM relative to 2μm away saturates at a value of about -1150 kHz at a height of 1nm above the sample, where the nominal range of sheet resistance values of the sample are 15Ω to 150Ω. Without the protrusion, the frequency shift would have followed the logarithmic dependence and reached a value of about -1500 kHz.

78.70.Gq, 84.37.+q, 07.79.Cz, 78.40.Fy

---

[1] Email: aimtiaz@squid.umd.edu

**Introduction:**
Near-Field microwave and infrared techniques have proven useful for extracting materials properties on the surfaces of many condensed matter systems for both fundamental and applied physics[1,2]. For example, on the fundamental physics side they have been used for quantitative imaging of dielectric permittivity, tunability, and ferroelectric polarization of thin dielectric films[3]. On the applications side they have been used to study electromagnetic fields in the vicinity of active microwave devices[4]. In Near-Field measurements, one often illuminates a probe with electromagnetic waves when the probe is held close to the surface of the sample. This probe to sample separation (d) is much less than the wavelength ($\lambda$) of the incident radiation. As a result the electromagnetic fields are highly confined in space and are very sensitive to probe-sample separation. Due to this extreme sensitivity, sample topography and probe geometry often become convolved with the material properties of the sample. For most purposes, the information of interest is the materials property. In order to de-convolve the required information, it is important that probe geometry and topography effects are understood, so their effects can be eliminated.

The next important frontier of near-field measurements is to achieve nanometer (nm) resolution imaging of materials properties. One way of achieving high spatial resolution is to bring the probe to a height equal to the size of a few unit cells. This high spatial resolution is possible due to the extreme sensitivity of the electromagnetic fields on height. We have achieved near-field electromagnetic microscopy by integrating a Scanning Tunneling Microscope (STM) feedback circuit with a NSMM; described in detail elsewhere[5]. This is similar to other near field microscopes utilizing STM to concentrate RF and microwave electric fields[6-10]. We have shown that our microscope is sensitive to the capacitance between the probe and sample as well as to the material properties of the sample.

The microwave microscope consists of a coaxial transmission line resonator coupled via a capacitor to the microwave source and feedback circuit[5]. The feedback circuit keeps track of the changes in resonant frequency of the resonator (producing a frequency shift signal, $\Delta f$) and the Quality factor (Q) of the resonator. The probe end of the resonator has a sharp metal tip sticking out of it, and this same tip is used to do STM and microwave microscopy (see inset of Fig.1). A simple lumped element model for the probe to sample interaction is a capacitor $C_x$ (typical value ~10 fF at a height of 1 nm) in series with a resistor $R_x$ in the case of conducting samples. The capacitance between the outer conductor and sample is so large that it presents a negligible impedance. The resistor contains the materials information of interest in this case. One can employ a transmission line model for the microscope[1], in which case the sample appears as a complex impedance given by $Z_x = R_x + \frac{1}{i\omega C_x}$. In the limit $\omega C_x R_x \ll 1$, the resulting frequency shift ($\Delta f$) is independent of $R_x$ and is directly related to $C_x$ as

$$\frac{\Delta f}{f_0} = \frac{-C_x}{2C_0}$$

, where $f_0$ is the resonant frequency of the unperturbed resonator and $C_0$ is the total capacitance of the microscope. In this case the microscope quality factor Q contains the loss information of the sample. The validity of this simple lumped-element treatment of the sample has been established for conducting films with low sheet resistance, i.e. in the limit $\omega C_x R_x \ll 1$[11].

The probes we used are commercially available Silver coated Tungsten tips. We find such tips to be good for both STM and NSMM. The measurement was performed on a variably Boron doped n-type Silicon sample, doped with the Focused Ion Beam (FIB) technique[12]. The energy of the Boron beam was 30 keV, and it was used to dope a total area of 10 μm x 10 μm. At this beam energy the depth of Boron implantation into Silicon is approximately 100 nm[13]. After ion-beam deposition the sample was rapid thermal annealed (RTA) to 900°C in Nitrogen for 20 seconds (5 seconds to ramp up the temperature, 10 seconds to anneal at 900°C and then 5 seconds to ramp down the temperature) to activate the carriers. This produced a roughly uniform dopant concentration over a thickness t = 100nm. The area is doped in stripes with the same concentration along the 10μm length, and varying concentration in the perpendicular direction. The concentration varies from $10^{16}$ ions/cm$^3$ to $10^{21}$ ions/cm$^3$ (resistivity $\rho = 10^{-2}$ Ω.m to $10^{-6}$ Ω.m) across this width, and there is a one to one relationship between concentration and resistivity for Boron doping in Silicon[14]. In this geometry, the resistivity can be converted to sheet resistance given by $R_x = \rho/t$ (one can also write $R_x = \frac{1}{q\mu p}$ where q is the magnitude of the electron charge, μ is mobility of holes and p is dose of Boron in ions/cm$^2$) and the Silicon substrate can be treated as a dielectric. This thin film model works because the undoped Silicon substrate has resistivity 0.61Ω.m, which is much larger than the doped film. The utility of this sample is that the surface is topography-free and the only variations in the sample are due to sheet resistance ($R_x$). This leaves our microwave signal mainly sensitive to material properties and probe geometry. The results reported in this paper satisfy the condition $\omega C_x R_x \ll 1$, since $R_x$ is in the range of 15Ω to 150Ω, ω is 2π 7.47 x $10^9$ Hz, and $C_x$ is on the scale of $10^{-14}$ F. This gives the additional benefit that the microscope is sensitive mainly to the probe geometry, since $\Delta f$ is independent of $R_x$, as mentioned earlier.

As mentioned earlier, for nanometer spatial resolution, the height of the probe above the sample should be on the nanometer length scale as well. With the help of a scanning tunneling microscope (STM) feedback circuit, we are able to reach and maintain a nominal separation of 1 nm above the sample. We are able to study the detailed dependence of frequency shift versus height in the range below 100 nm for the first time. As shown by the data of Fig.1, in this novel regime, we see a deviation from the expected logarithmic drop of frequency shift versus height (which would appear as a straight line in Fig.1). In this paper, we attempt to understand the origin of this deviation.

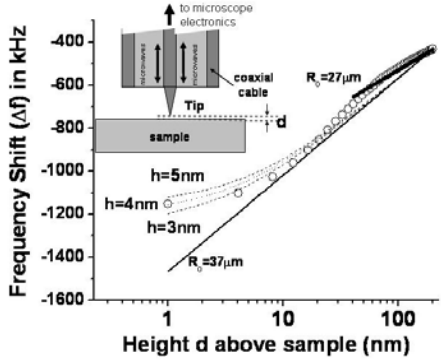

**Figure 1:** Frequency shift relative to 2μm height of a coaxial microwave microscope versus tip-sample separation d. The sample is a Boron doped Silicon substrate discussed in the text. The tip WRAP082 is used for measurements. The open circles show frequency shift (Δf) data; Solid lines are Δf values calculated using capacitance values from the simple sphere above the plane model; dashed lines are Δf values obtained from calculated capacitance values with a small cone added to a sphere of radius $R_0 = 37$μm. Inset shows a schematic cross-section diagram of the very end of the microscope interacting with the sample (not to scale).

To perform the STM, the bias on the sample was +1 volts and the tunnel current set point for STM feedback was 0.5nA. Such high bias is needed in order to overcome the 0.7 volts of voltage drop across the buried p-n junction of the Boron-doped Silicon. The microwave source frequency was 7.472 GHz.

**Model:**
As mentioned earlier, experimentally, the very end of the probe is coaxial with the inner conductor protruding beyond the end of the coaxial cable (inset of Figure 1). The protruding section of the inner conductor has a roughly conical geometry. The same tip is used for STM and electromagnetic wave illumination and concentration. To determine the capacitance between the probe and sample, one can model a conical tip sticking out of the inner conductor of the coax, and calculate its capacitance with respect to a conducting plane. On millimeter and micron length scales, the outer conductor of the coax has to be taken into account in order to correctly calculate the probe-sample capacitance. However, our interests are for measurements in the sub-micron height regime, where the outer conductor capacitance does not change significantly. One option is to model the tip as an infinite cone. However, an infinite cone above a sample is a scale invariant problem, as it puts no length scale for the probe itself. We will model the probe as a sphere above a sample, where the sample is an infinite plane[15]. Such a model gives another length scale, which is the sphere radius $R_0$, in addition to the height (d) of the probe (measured from the bottom of the sphere to the flat sample) above the infinite plane.

This model has been applied earlier to study the behavior of microwave microscopes on sub-micron length scales, however, far away from the atomic length. The strength of this model lies in analytic expressions for capacitance and static electric field obtained through an image charge method.

This model has also been successfully extended to understand the dynamics of electromagnetic interactions between probe and sample[15]. However, in this paper we limit ourselves to the static aspects.

Electromagnetically speaking, the tip to sample interaction in our microwave microscope is very complex, and modeling it analytically alone is a very challenging task. We resort to numerical techniques when analytic models become difficult. For example, the addition of even a small object at the end of the probe makes the analytic problem very complex to solve. The numerical modeling was performed with commercially available Ansoft Maxwell 2D (M2D) software, which solves a variety of static electric and magnetic problems. We model the measured nm-scale deviations by adding a small geometrical object (a tiny cone) at the end of the sphere. The height of cone is 'h' and the cone and sphere together act as a probe above the sample.

**Results:**
Our goal is to understand the Δf versus height above the sample data for heights below 10 nm, as shown in Fig.1. For heights above 10 nm, the interpretation is simple as one gets close to the sample. The effective area of the parallel plate capacitor is changing at the same time as the distance d is changing. In Fig.1, we show this by two solid lines for the analytic sphere above the plane model. A sphere of $R_0 = 27$μm describes the data at large heights (d > 50 nm), and a sphere of radius $R_0 = 37$μm works well as one gets closer to the sample. However this trend changes below 10 nm where we see saturation in the Δf signal and this saturation is seen above all samples measured so far. What follows here is a systematic discussion to understand this saturation.

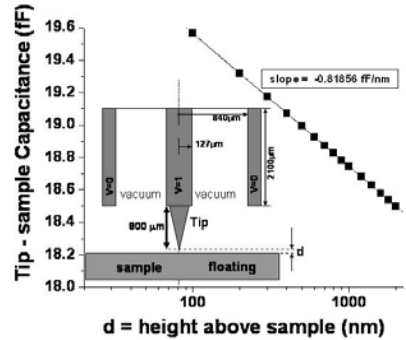

**Figure 2:** Tip to sample capacitance from Maxwell 2D (M2D) for the coaxial probe geometry shown in cross-section in the inset. The outer conductor is included in the model and the different voltages (V in volts) and lengths are shown on the drawing in the inset (not to scale). The plot shows the logarithmic drop of capacitance with increasing d calculated from this model. The tip is conical except for last 14.2μm which is an ellipsoid with a flat bottom facing the sample. The box size for the problem was 2500μm by 2100μm.

First we start with a model which is closest to the experiment. Fig.2 is a plot of capacitance versus height for a coaxial probe with conical tip above a conducting sample, determined numerically using Maxwell 2D. The geometrical model is shown in cross-section in the inset of Fig.2. The center conductor was set at a potential of 1 Volt, the outer conductor was set to 0 Volts, and the sample was left to float.

Here we use the approximation that the near-field structure will be dictated by the static electric field structure. This approximation is valid, since in our case the source (probe tip) and sample separation satisfy the condition d << λ; where we assume that electromagnetic sources and fields have harmonic ($e^{-i\omega t}$) dependence. With this assumption one can write down the equations for the potentials and the fields as a function of position and time. After integration over time, the final solution for fields in the "near- zone" (d<<λ) shows the quasi-static nature of the fields[16].

The key point to notice in Fig. 2 is that the probe to sample capacitance for a conical tip has a logarithmic increase as a function of decreasing height, even in the sub-micron regime. Any model that we use should maintain this qualitative behavior. Due to aspect ratio limitations of the M2D software (ratio of largest length scale to smallest must be less than ~10³), we were not able to push this study to much smaller separation, d. However, the sphere above a semi-infinite plane model can be taken to these small length scale values in M2D, since $R_0$ is much smaller than the dimensions of the full coaxial probe and we have more room to work around aspect ratio problems. However, before we discuss that, it is better to examine the geometry of the tips used.

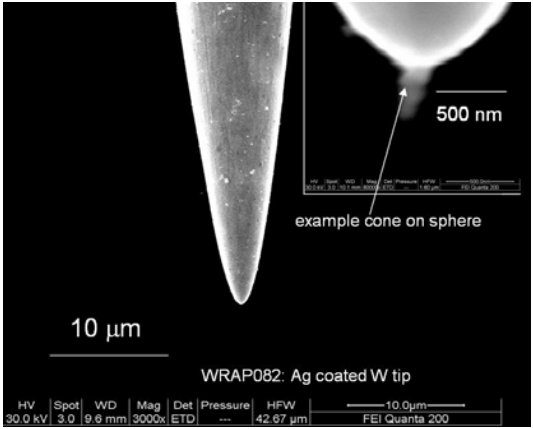

**Figure 3: Scanning Electron Microscope micrograph for one of the STM probe tips, WRAP082. Inset shows the magnification of the apex of the tip, which illustrates the cone shaped "particle" sticking to the probe. This tip was imaged before use in the NSMM.**

*Geometry of the probes:*
The metal tips in our experiment are first imaged by a Scanning Electron Microscope (SEM), where they reveal a rounded conical nature (Fig.3). On the millimeter and micrometer length scale, the probe can be well approximated as a cone. Near the tip, the cone is rounded and includes a sphere of radius r. Note the protrusion sticking out of the tip, shown in the inset of Fig.3. This small protrusion is probably left from the tip preparation procedure. The size of protrusion at the apex of the tip is few hundred nanometers (see the inset of Fig.3) in this case. In general, the tip can pick up material from the sample, and there can be other sources of damage. Protrusions like this can have a significant effect on the measurement.

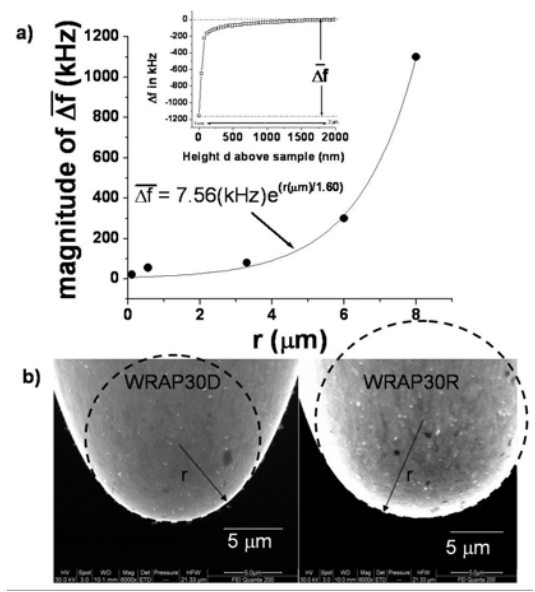

**Figure 4: a) magnitude of the frequency shift of the NSMM between tunneling height and 2 μm (defined as $\overline{\Delta f}$ ) on a gold/mica thin film sample, versus inscribed sphere radius r. Note that $\overline{\Delta f}$ increases roughly exponentially with increasing r for different probes (inset shows the definition of magnitude of $\overline{\Delta f}$ ); b) SEM micrograph showing r for two different probes WRAP30D (r≈6μm) and WRAP30R (r≈8μm).**

A second observation is that the tip-sample capacitance (as measured through Δf) is proportional to the sphere radius r. On the sub-micron scale one can easily see an effective sphere with radius 'r' (this r is distinct from R0) embedded at the end of the probe (which is illustrated in Fig.4b). For the probes shown in the figure the approximate r value is 6μm (WRAP30D) and 8μm (WRAP30R). Fig.4(a) shows that as r gets larger, the magnitude of frequency shift signal between tunneling height and 2 μm (defined as $\overline{\Delta f}$ ) gets larger. Here the sample is a gold thin film on mica satisfying $\omega C_x R_x$ << 1, so in this case the total frequency shift $\overline{\Delta f}$ is directly proportional to capacitance change. We arbitrarily define the zero of $\overline{\Delta f}$ to be the height of 2μm. These data illustrate the dramatic dependence of the frequency shift on probe dimension, and help justify modeling of the probe as a sphere, at least on sub-micron length scales. However, the tips of large r, which show this dramatic effect for frequency shift signal, are very poor for STM. These tips enlarge the topographic features laterally (which can happen if the tip has multiple tunneling sites) and give false heights of the features as well.

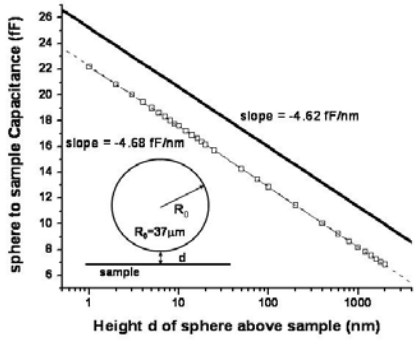

**Figure 5:** Comparison of sphere to sample capacitance (inset shows the geometry) calculated from the analytic model (solid line) and M2D (data points). The sample is infinite in the analytic case. The sphere radius $R_0 = 37$ μm. The box size for M2D is 90μm by 91μm. The boundary condition on the box was Neumann (electric field tangent to the boundary of the box). The slopes agree to within 1.3%.

*Discussion of results for sphere above plane:*
Figure 5 is a plot of the capacitance of a sphere above a semi-infinite plane calculated two different ways (analytic and numerical), as the height is varied. It also clearly shows the logarithmic increase of capacitance versus decreasing height, for both the analytical model (solid line) and numerical model (points). This simple model thus preserves the logarithmic dependence noticed in Fig.2. The slopes are very close to each other (differing by only 1.3%). The offset of the numerical curve is due to the limited size of the numerical simulation. Figure 6 shows the numerically determined capacitance versus the inverse of the numerical problem size (area of drawing in M2D, called the box size) for different fixed heights above a conducting sample. It shows that in the limit of infinite box size, the extrapolated numerical result agrees with the analytic result to within 1.4%. For example for the height of 1nm, the analytic value of capacitance is 25.58fF. In the limit of infinite box size, M2D yields value of 25.23fF (the graph also shows the comparison for two other heights).

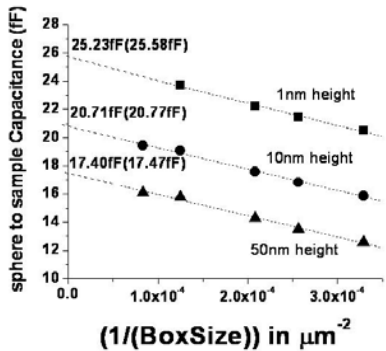

**Figure 6:** Comparison of sphere to sample capacitance as a function of inverse box size of the M2D calculation. The numbers in parentheses show the projected value of capacitance that M2D would yield in the limit of infinite box size (i.e. the y-intercept of the straight line fits). The extrapolated values are within 1.4% of the analytic values at the same height.

We note that the sphere radius $R_0$ required to fit the $\Delta f(d)$ data is much larger than the embedded sphere radius r. This is because the embedded sphere intercepts only a fraction of the electric field lines between the tip and the sample, thus missing the field lines between the rest of the non-spherical probe and sample. The net effect of all these field lines is incorporated in an artificially enhanced $R_0$ for the model sphere. We find that the value of $R_0 = 37$μm fits our data best for heights down to 10nm, for the tip shown in Fig.1.

*Discussion of results with small cone added to sphere above the plane:*
In order to study the deviation of our tips from spherical shape, we added a small cone at the end of the sphere (inset of Fig.7). This is motivated partly by the cone shaped "particle" seen in the SEM micrograph (shown in the inset of Fig.3). We find (numerically) that other shapes produce similar results. This new cone has two lengths associated with it, namely the base and height of the cone. We found that the height of this cone is the only key parameter affecting the overall capacitance, so we took the base to be fixed (2nm) for all the heights. This model was studied numerically, since the analytic problem with this perturbation (cone) is difficult to solve. As shown in Fig.7, adding this perturbation produces a deviation from the logarithmic increase of capacitance with decreasing height. The effect on the capacitance sets in at heights on the same length scale as the size of the cone, and it tends to saturate the capacitance. The larger the value of h, the more prominent is the saturation effect.

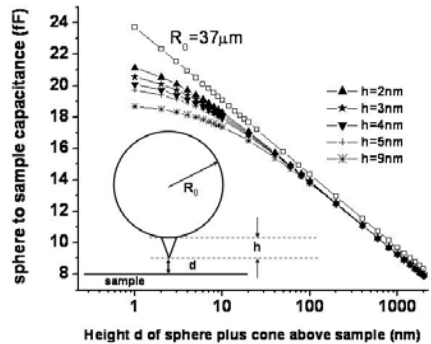

**Figure 7:** Sphere (with radius of $R_0 = 37$μm) to sample capacitance with a small cone of height h added at the end of the sphere using M2D (inset shows the geometry, which is not to scale). The capacitance saturates at heights d on the same scale as h. Plot shows results for several different values of the cone size h. The box size is 80μm by 81μm. The boundary condition on the box was Neumann (electric field tangent to the boundary of the box).

In Fig.1, the dotted curves show these capacitance values converted into frequency shift using a transmission line model of the microscope6. The frequency shift data is plotted as open dots for comparison. We find that a sphere of radius $R_0 = 37$μm with a perturbation cone height of 3nm to 5nm fits the data well.

**Conclusion:**

We report a new length scale in tip-to-sample interaction for Near-Field microwave measurements. The origin of this length scale is a deviation of the probe geometry from an ideal (spherical or conical) geometry. This length scale is typically on the order of 3nm to 5nm and will effect measurements on these length scales. Such results are important to design an optimum probe for the future development of near-field microwave microscopes.

**Acknowledgements:**
This work has been supported by an NSF Instrumentation for Materials Research Grant DMR-9802756, the University of Maryland/Rutgers NSF-MRSEC and its Near Field Microwave Microscope Shared Experimental Facility under Grant number DMR-00-80008, the Maryland Industrial Partnerships Program 990517-7709, the Maryland Center for Superconductivity Research and by a Neocera subcontract on NIST-ATP# 70NANB2H3005. We also thank Neocera for allowing us to use their SEM facility, and Nolan A. Ballew for doing the rapid thermal anneal (RTA) on the sample.